# Time-resolved vacuum Rabi oscillations in a quantum dot-nanocavity system

K. Kuruma[1], Y. Ota[2], M. Kakuda[2], S. Iwamoto[1,2] and Y. Arakawa[1,2]

[1]*Institute of Industrial Science, The University of Tokyo, 4-6-1 Komaba, Meguro-ku, Tokyo 153-8904, Japan*

[2]*Institute of Nano Quantum Information Electronics, The University of Tokyo, 4-6-1 Komaba, Meguro-ku, Tokyo 153-8904, Japan*

We report time-domain observation of vacuum Rabi oscillations in a single quantum dot strongly coupled to a nanocavity under incoherent optical carrier injection. We realize a photonic crystal nanocavity with a very high quality factor of >80,000 and employ it to clearly resolve the ultrafast vacuum Rabi oscillations by simple photoluminescence-based experiments. We found that the time-domain vacuum Rabi oscillations were largely modified when changing the pump wavelength and intensity, even when marginal changes were detected in the corresponding photoluminescence spectra. We analyze the measured time-domain oscillations by fitting to simulation curves obtained with a cavity quantum electrodynamics model. The observed modifications of the oscillation curves were mainly induced by the change in the carrier capture and dephasing dynamics in the quantum dot, as well as the change in bare-cavity emission. This result suggests that vacuum Rabi oscillations can be utilized as a highly sensitive probe for the quantum dot dynamics. Our work points out a powerful alternative to conventional spectral-domain measurements for a deeper understanding of the vacuum Rabi dynamics in quantum dot-based cavity quantum electrodynamics systems.

# I. INTRODUCTION

Semiconductor nanostructures coupled to optical resonators are a fascinating platform for studying solid-state cavity quantum electrodynamics (CQED)[1,2]. In particular, CQED systems based on quantum dots (QDs) and photonic crystal (PhC) nanocavities are some of the most advanced systems, due to their strong light-matter interactions originating from the tight optical confinement of the nanocavities both in time and space. Since the first demonstration of strong coupling in QD-CQED[3,4], the coherent light matter interactions have been employed for developing various classical and quantum optical devices including optical switches[5–7], non-classical light generators[8,9] and quantum gates[10].

Meanwhile, tremendous effort has been devoted to understanding the physics of strongly coupled QD-CQED systems. Most of such studies were carried out by measuring photoluminescence (PL) spectra under incoherent optical carrier injection. The coupling strength in a QD-CQED system can be quickly characterized by measuring its vacuum Rabi splitting[3,4,11–13]. Moreover, the PL-based experiments have been widely employed for studying various intriguing phenomena such as triplet vacuum Rabi spectra[14–17], off-resonant cavity-QD coupling[14,18–21], phonon-QD-cavity coupling[22,23] and pump-induced dephasing[24]. However, the spectral domain approach often encounters the difficulty in resolving the effect of slow dynamics, which only slightly modifies the spectrum and is likely to be overlooked due to the limitation in spectrometer resolution.

For investigating the dynamics in strongly-coupled QD-CQED systems, time domain spectroscopy is the most straightforward approach. Previous studies have employed ultrafast spectroscopy techniques using cavity-resonant optical pulses and succeeded in resolving vacuum Rabi oscillations in QD-CQED systems[25,26]. However, the observed oscillations were highly dissipative due to fast photon leakage from the resonators and were recorded with rather limited signal to noise ratios, making them unsuitable for analyzing slow dynamics in the CQED systems.

In this work, we report time domain observation of vacuum Rabi oscillations in a QD strongly coupled to a PhC nanocavity under optical carrier injection. We realized a PhC nanocavity with a very high experimental quality ($Q$) factor of > 80,000, enabling the observation of remarkably coherent vacuum Rabi oscillations by conventional time-correlated single photon counting (TCSPC). We utilized the observed vacuum Rabi oscillations for probing the inner workings of the QD-CQED system in conjunction with curve fitting using a theoretical CQED model. The sensitive responses of vacuum

Rabi oscillations to the QD dynamics facilitate the detection of slow pump-induced dephasing in the QD, which was not resolved by our spectrometer. We also observed that the carrier capture process in the QD largely depends on pump laser wavelength. We believe that our approach based on PL spectroscopy will advance the understanding of the dynamics in QD-CQED systems driven by carrier injection and thus be of importance for various electrically-pumped QD-based CQED devices[27–31].

## II. SAMPLE STRUCTURE AND EXPERIMENTAL SETUP

In this study, we used a GaAs-based PhC double-heterostructure cavity[32] with a lattice constant of $a_1 = 252$ nm and an air-hole radius of $r = 61$ nm. The lattice constant of the central hetero region ($a_2$) is slightly elongated to be 259.6 nm along the waveguide, forming an optical mode gap to support nanocavity modes. This nanocavity design enables the realization of high experimental $Q$ factors[33], which are advantageous for observing time-domain vacuum Rabi oscillations. For better vertical extraction of the cavity emission, we applied double-periodic modulations of airhole radii around the cavity region[34]. Figure 1(a) exhibits an electric field profile for the fundamental cavity mode calculated by a finite-difference time-domain algorithm. This cavity mode confines light with a $Q$ factor of ~$3 \times 10^5$ and a mode volume of $1.5(\lambda/n)^3$, where $\lambda$ (= $3.875a$) and $n$ (= $3.46$) are the cavity resonant wavelength and the refractive index of GaAs, respectively. The designed cavity was fabricated into a 130 nm-thick GaAs slab on top of a 1-µm-thick $Al_{0.7}Ga_{0.3}As$ sacrificial layer, which is later dissolved to form an airbridge structure. The details of the nanofabrication processes can be found in our previous report[35]. The GaAs slab contains a single layer of InAs QDs in the middle. The QD density was estimated to be ~$10^8$ cm$^{-2}$. Figure 1(b) shows a scanning electron microscope (SEM) image of a fabricated cavity.

For optical characterizations, we performed micro photoluminescence (µPL) measurements in both spectral and time domain using an optical setup illustrated schematically in Fig. 1(b). The PhC sample was kept in a helium flow cryostat at cryogenic temperatures (4-10 K). We used an 808-nm continuous wave (CW) diode laser for the above-bandgap excitation of the GaAs barrier around the InAs QDs. We also employed a CW wavelength-tunable laser for PL excitation (PLE) measurements. For time domain measurements, we employed a Ti:sapphire pulse laser with a pulse duration of 1 ps and a repetition rate of 80 MHz. We focused the pump laser light onto the PhC cavities by an objective lens (OL) with a numerical aperture of 0.65. PL signals from the sample were spectrally resolved by a spectrometer and subsequently detected with a Si CCD camera. The spectrometer resolution was experimentally measured to be 21 $\mu$eV. During PL measurements, we precisely controlled the cavity resonant energy using a Xe gas condensation technique[36]. For time-resolved PL experiments, we used a TCSPC (Becker & Hickl Corp.) system equipped with a fast-response superconducting single photon detector (SSPD, SCONTEL Corp.). We note that the SSPD supports a high detection efficiency of ≥25% with a very low dark count rate of ≤10 counts/sec in the investigated wavelength region (~1$\mu$m), which are highly beneficial for time-resolved PL experiments with better signal to noise ratios. We used the spectrometer as a spectral bandpass filter (bandwidth = 180 $\mu$eV) for PL signals before being sent to the SSPD. The total time resolution of the setup was measured to be 25.6 ps, which is mainly determined by the timing jitter of our SSPD.

**III. BASIC OPTICAL CHARACTERIZATION**

First, we characterized the target QD-cavity system at 10 K. Figure 1(c) shows a bare cavity spectrum measured under an excitation power of 7 nW (measured before the OL). By fitting the spectrum with a Voigt peak function with a fixed Gaussian part representing the spectrometer response, we obtained a high cavity $Q$ factor of 81,000 (corresponding to a cavity decay rate $\kappa$ = 16 $\mu$eV). This $Q$ factor is, to the best of our knowledge, the highest value reported to date for any PhC nanocavity employed for QD-CQED studies. Note that the excitation power used for measuring the cavity PL spectrum was so weak that the estimated $Q$ factor can be regarded as the intrinsic $Q$ factor unaffected by the carrier pumping process. We did not find any significant change in the cavity $Q$ factor during the cavity frequency tuning by the gas condensation.

Next, we investigated the optical coupling properties between the cavity mode and a QD exciton transition peak at 1.31395 eV. Figure 2(a) shows a summary of PL spectra taken by tuning the cavity mode resonance across the QD transition. Near the QD-cavity resonance, an anti-crossing behavior was observed, suggesting that the QD-cavity system is in strong coupling regime. From the peak split, we deduced a vacuum Rabi splitting of 35 $\mu$eV, which can be translated into a coupling strength (*g*) of 18 $\mu$eV. Figure 2(b) shows an on-resonance spectrum, composed of two polariton peaks and an additional center peak. The center peak is of the bare cavity mode emission[14], which occurs when transition peaks of the QD are off resonant from the cavity resonance and is likely to be predominantly supplied by unspecified background emission in the sample[16]. When fitting to the vacuum Rabi spectrum, we fixed the linewidth of the center peak to that of the bare cavity mode emission recorded under the far detuned condition.

For characterizing absorption levels in the investigated QD, we performed PLE measurements[37] at 7 K by using the tunable laser with an excitation power of 160 μW. Figure 2(c) and (d) respectively show a PL spectrum of the QD emission and a corresponding PLE spectrum measured with various excitation laser detunings spanning from 19.4 to 63.8 meV. In the PLE spectrum, we observed several sharp resonances together with a broad background across the measurement range. We attributed the sharp peaks to optical transitions involving higher energy levels in the QD and those assisted by optical phonons. We consider that the continuous background predominantly originates from the absorption tail of a wetting layer (WL) contacting with the QD.

## IV. TIME-DOMAIN OBSERVATION OF VACUUM RABI OSCILLATIONS

### A. Detuning dependence

For the observation of vacuum Rabi oscillations, we carried out time-resolved PL measurements at 6 K. We set the laser center energy to 1.36564 eV (detuned from the QD by 52 meV), which primarily resonates with one of the PLE peaks in the QD. The average excitation power was fixed to be 160 μW. Figure 3(a) shows time-resolved PL spectra observed for three different spectral detunings ($\delta$ = cavity mode detuning from the QD transition). Close to the zero detuning condition ($\delta$~0), we observed a decaying curve with clear intensity oscillations with a time period of ~117 ps (extracted by the Fourier transform of the oscillating curve). This period agrees well with that expected from the coupling constant ($2g$ ~ 118 ps) deduced from the spectral measurements. When increasing $|\delta|$, we observed a decrease of the time period of oscillations, together with the decrease of peak-to-peak oscillation amplitudes. Figure 3(b) shows a summary of the measured oscillation time periods as a function of $\delta$. The periods show a hyperbolic dependence on $\delta$ and are well comparable with a simplified expression of the Rabi frequency, $\Omega \sim \sqrt{g^2 + \delta^2/4}$ (solid line). Here, we neglected the influence of damping terms of $\kappa$ and QD spontaneous emission decay ($\gamma$). These observations firmly demonstrate the time-domain detection of vacuum Rabi oscillations under optical carrier injection.

We fit the measured PL decay curves with simulations based on a CQED model as plotted with solid lines in Fig. 3(a) (see Appendix for the fitting process details). In the model, we considered experimentally measured parameters ($\kappa$ = 16 μeV and $g$ = 18 μeV) and a few other fitting variables. We included the contribution from the bare cavity emission, which originates from the additional center peak in the vacuum Rabi spectra and radiates with a measured PL decay time of $T$ ~360 ps. This decay constant is determined by the average photon supply rate from background emitters enclosed within the nanocavity. We set the amount of the bare cavity emission ($A_i$) as a fitting parameter. The incorporation of the bare cavity component was found to be essential to reproduce the slowly decaying tails in the measured curves (prominent for the cases of small $\delta$s). We also took QD's pure dephasing rate ($\gamma_{ph}$) and the carrier relaxation rate from a QD excited state to its radiative state ($\gamma_R$) as fitting parameters. All the measured decay curves are well reproduced by calculating the time-resolved cavity radiation intensities with a single quantum master equation incorporating the CQED model.

**B. Excitation wavelength dependence**

Next, we investigated the excitation wavelength dependence of vacuum Rabi oscillations under the QD-cavity resonance at 4 K. We scanned the center energy of the pulsed laser source from 1.53465 eV (above the GaAs band gap) to 1.33621 eV (21 meV above the QD exciton line). In the current measurements performed in a later day than that for Fig. 3, the cavity $Q$ factor was slightly degraded to be 71,000 ($\kappa = 19\ \mu$eV), probably due to unspecified surface contamination. Figure 4(a) shows time-resolved PL spectra measured with different laser center frequencies. For each point of the measurements, we varied pump laser power so as to keep the same photon count rate on the SSPD. We observed clear modifications of the measured time-resolved curves: the Rabi oscillations become clearer with decreasing excitation energy. Concomitantly, the measured curves become more likely to decay faster, suggesting a reduction of the slow decay component originating from the bare cavity mode emission. We also found faster initial rises of the emission intensity when using lower energy excitations, as summarized in Fig. 4(b). In contrast, despite these large modifications in the time resolved spectra, we did not observe significant changes in the corresponding PL spectra, as plotted in Fig. 4(c). This comparison accentuates the usefulness of the time-domain spectroscopy for studying QD-based CQED.

We analyzed the measured PL decay curves by comparing with the CQED model in terms of its main fitting parameters ($\gamma_R$, $A_i$, $\gamma_{ph}$). In Fig. 4(a), we overlaid the fitting curves (solid lines) on the measured curves. Through the fitting process, we deduced the values for $\gamma_R$, $A_i$ and $\gamma_{ph}$, which are respectively summarized in Figs. 4(d), (e) and (f). The values of $A_i$ are normalized to that measured by the above bandgap excitation. In Fig. 4(d), we found a tendency of faster carrier relaxation for lower excitation energies, which can also be confirmed by the sharper initial intensity rises in Fig. 4(b). A steep increase of $\gamma_R$ occurs when the laser detuning becomes less than 70 meV, roughly corresponding to the lower energy edge of the WL absorption peak. Under such low energy excitations, carriers are tend to be directly injected into the QD, eliminating the time of carrier diffusion in the surrounding material before the carrier capture and hence resulting in a faster $\gamma_R$. The monotonic increase of $\gamma_R$s with decreasing laser energy may be understood as a result of a gradual increase of the contribution of the faster direct carrier capture process.

We also observed a monotonic decrease of $A_i$ when reducing laser detunings below 70 meV, as shown in Fig. 4(e). It is known that excess carriers within the barrier material can turn into background emission uncorrelated with the QD, which, we assume, is a primary source for the bare cavity mode emission. Supposing $A_i$ as an indicator of excess carriers in the surrounding material, its reduction for lower laser frequencies consistently explains the accompanied increase of $\gamma_R$ by the reduced indirect carrier capture processes through the surroundings. The evolution of deduced $\gamma_{ph}$s is summarized in Fig. 4(f), showing a nearly-monotonic increase for higher laser frequencies. Again, this observation can be attributed to an increase of the excess carriers within the surrounding material, the fluctuation of which is known to induce pure dephasing in the QD[38,39]. It is noteworthy that we extracted $\gamma_{ph}$ of only a few $\mu$eV by fitting to the time domain vacuum Rabi oscillations. Such small pure dephasing is in general difficult to resolve by spectral domain PL measurements, highlighting the usefulness of the time-domain approach for understanding the slow dynamics in QD-CQED.

**C. Excitation power dependence**

Finally, we investigated the excitation power dependence of the vacuum Rabi oscillation under the QD-cavity resonance with 1.33909 eV pumping (25 meV above the QD transition). Figure 5(a) displays time-resolved PL spectra measured for three different excitation powers of 10, 113 and 300 $\mu$W. As the pump power increases, the Rabi oscillation becomes blurred and the time resolved PL curve acquires more slow decay component. The increase of the slow component is again attributed to an increase of the bare cavity mode emission, which is also indicated by the slightly reduced center dips in the corresponding vacuum Rabi spectra shown in Fig. 5(b).

Using the same fitting procedure, we again estimated the values of the main fitting parameters ($\gamma_R$, $A_i$, $\gamma_{ph}$) for different pump powers. Figure 5(c) shows the deduced $\gamma_R$s as a function of average pump power. $\gamma_R$s are found to be nearly constant around a very fast value of ~38 $\mu$eV (58 GHz), which can be thought of as a result of the dominant direct carrier injection into the QD. Meanwhile, as seen in Fig. 5(d), we observed a clear increase of the bare cavity emission $A_i$ for stronger laser pump, suggesting an increase of excess carriers in the surroundings. We consider that, in the current experiments, the laser energy is very low and thus the excess carriers are mainly generated in deep trap levels of the barrier materials. As such, they cannot efficiently diffuse to be captured by the QD and thereby do not significantly alter $\gamma_R$. Consistent with the increase of $A_i$, we observed an increase of $\gamma_{ph}$ from 2.6 $\mu$eV (3.9 GHz) to 6.4 $\mu$eV (9.7 GHz) when using stronger pump powers, as depicted in Fig. 5(e). We consider that the monotonic increase of $\gamma_{ph}$ directly reflects the effect of pump-induced dephasing[24,38,39]. Again, the extracted rate of the pump-induced dephasing at minimum is only a few $\mu$eV, which is much smaller than our spectrometer resolution, suggesting the advantages of time-domain measurements for discussing the slow QD dynamics.

## V. CONCLUSION

In conclusion, we demonstrated the time-domain observation of vacuum Rabi oscillations in a strongly coupled QD-PhC cavity system driven under incoherent optical carrier injection. We utilized a high $Q$ (=81,000) PhC nanocavity and observed clear vacuum Rabi oscillations by simple PL-based experiments using a TCSPC system. We further performed the time-domain experiments with varying the excitation wavelength and power. We found that the vacuum Rabi oscillation profiles largely change even when the corresponding PL spectra did not change significantly. We analyzed the measured oscillation curves by fitting to a theoretical CQED model. We concluded that the observed modifications in the time-domain oscillation curves originate from the dephasing and carrier capture process in the QD, as well as change in the bare cavity emission intensity. These findings suggest that the vacuum Rabi oscillations can be used for a highly sensitive probe for the dynamics in QDs. Our study also will be a great help for quantum optical applications based on vacuum Rabi oscillations[40] as well as for QD-CQED devices driven by carrier injection .

**ACKNOWLEDGMENTS**

We would like to thank M. Nishioka, S. Ishida and C. F. Fong for their technical support and fruitful discussions. This work was supported by Japan society for the Promotion of Science (JSAP) Grants-in-Aid for Specially Promoted Research (KAKENHI) No.

**APPENDIX: THEORETICAL MODEL**

We fitted the experimentally-observed vacuum Rabi oscillations using numerically-calculated PL decay curves. In the simulation model, we consider a three-level ladder-type quantum system for describing a QD, as illustrated in Fig. A1. The radiative transition between the middle ($|E\rangle$) and the lowest ($|G\rangle$) energy levels is assumed to coherently couple to a single cavity mode with a rate of $g$. The top energy level ($|U\rangle$) and its incoherent relaxation to $|E\rangle$ are introduced for discussing the carrier capture process. The Hamiltonian of this system under the dipole and rotating-wave approximation is given by

$$H = \sum_{i=U,E,G} \omega_i \sigma_{i,i} + \omega_c\, a^\dagger a + g(\sigma_{G,E} a^\dagger + H.c.) \quad (1)$$

$\omega_i$ and $\omega_c$ are the frequency of the QD state $|i\rangle$ and of the cavity mode, respectively. $\sigma_{i,j} = |i\rangle\langle j|$ is a pseudo spin operator, while $a^\dagger$ and $a$ are the creation and annihilation operator for cavity photons, respectively. In an appropriate rotating frame, the Hamiltonian can be rewritten as

$$H' = \delta\, a^\dagger a + g(\sigma_{G,E} a^\dagger + H.c.) \quad (2)$$

,where $\delta = \omega_c - (\omega_E - \omega_G)$. In the model, we also treat several incoherent processes including cavity photon leakage (at a rate of $\kappa$), emitter's spontaneous emission ($\gamma$), incoherent state relaxation from $|U\rangle$ to $|E\rangle$ ($\gamma_R$), incoherent state pumping from $|G\rangle$ to $|U\rangle$ ($P$) and emitter pure dephasing ($\gamma_{ph}$). Then, we obtain a quantum master equation as follows,

$$\frac{d\rho}{dt} = i[\rho, H'] + \mathcal{L}(\rho) \quad (3)$$

,where $\rho$ is the density operator for the QD-cavity system and $\mathcal{L}(\rho)$ is the Lindblad superoperator defined as:

$$\mathcal{L}(\rho) = \frac{\kappa}{2}(2a\rho a^\dagger - a^\dagger a\rho - \rho a^\dagger a) + \frac{\gamma}{2}(2\sigma_{G,E}\rho\sigma_{G,E}{}^\dagger - \sigma_{G,E}{}^\dagger\sigma_{G,E}\rho - \rho\sigma_{G,E}{}^\dagger\sigma_{G,E}) +$$

$$\frac{\gamma_R}{2}(2\sigma_{E,U}\rho\sigma_{E,U}{}^\dagger - \sigma_{E,U}{}^\dagger\sigma_{E,U}\rho - \rho\sigma_{E,U}{}^\dagger\sigma_{E,U}) + \frac{P}{2}(2\sigma_{G,U}{}^\dagger\rho\sigma_{G,U} - \sigma_{G,U}\sigma_{G,U}{}^\dagger\rho -$$

$$\rho\sigma_{G,U}\sigma_{G,U}{}^\dagger) + \frac{\gamma_{ph}}{2}(\sigma_z\rho\sigma_z - \rho) \quad (4)$$

where, $\sigma_z = |E\rangle\langle E| - |G\rangle\langle G|$. For the incoherent pumping, we introduced a Gaussian-shaped pulsed pumping: P = $P_0/(\tau\sqrt{2\pi})$exp(-$(t-t_0)^2/2\tau^2$). Here, $P_0$ describes the amplitude of the incoherent pumping, which was kept low enough for appropriately describing the actual experiments. $t_0$ and $\tau$ respectively determine the timing of the pulse irradiation and pulse duration. We set the full width at half maximum (FWHM) of the pulse $\tau_{\text{FWHM}}$ (= $\tau\times 2\sqrt{2\ln 2}$) to be 1 ps. We treated $g$, $\kappa$ and $\gamma$ as fixed parameters. We employed experimentally obtained values for the first two parameters. For $\gamma$, we used a typical value of 0.13 μeV for our QD-cavity systems[41]. We solved the master equation in the time domain based on the Runge–Kutta method, after truncating the cavity photon Hilbert space at the single photon Fock state. From the simulated density matrix, we calculated time evolutions of cavity photon number. We then added the contribution of slowly-decaying bare cavity mode emission expressed as a single exponential function: $A_i$exp(-$t/T$) + $y_0$. The resulting cavity emission curves were then used for fitting after being convolved with a peak function reflecting the detection system time response (25.6 ps). For the fitting, we treated the amplitude of the slow component ($A_i$) and the constant offset ($y_0$) as fitting parameters. The decay time constant ($T$) are measured to be ~360 ps by measuring cavity mode emission under a far off-resonant condition. All the fitting variables (Δ, $\gamma_R$, $\gamma_{ph}$, $A_i$ and $y_0$) are deduced by fitting the experimental data with theoretical curves based on a least squares method using a trust-region-reflective algorithm (included in the Optimization Toolbox, MATLAB, The MathWorks).

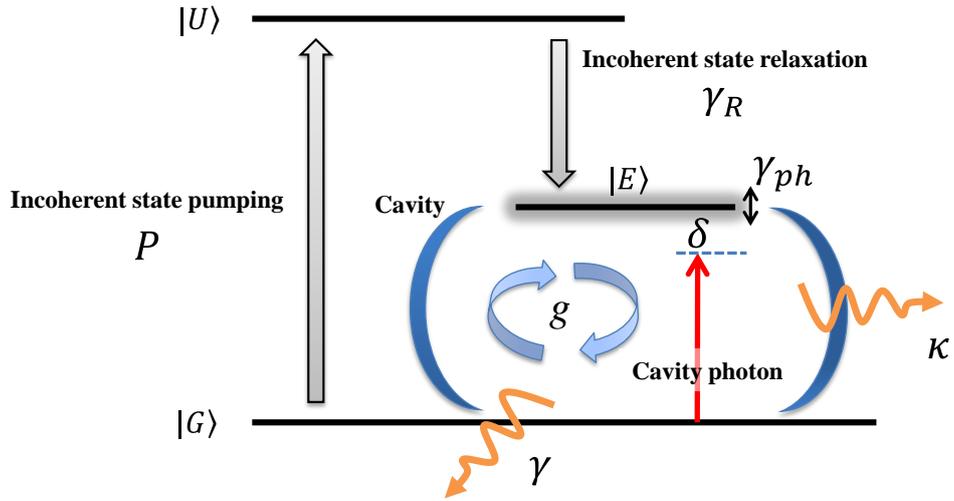

Fig. A1 Schematic illustration of the three-level system representing the QD. The middle ($|E\rangle$) and the lowest ($|G\rangle$) energy level are coupled to the cavity mode.

Figure 1

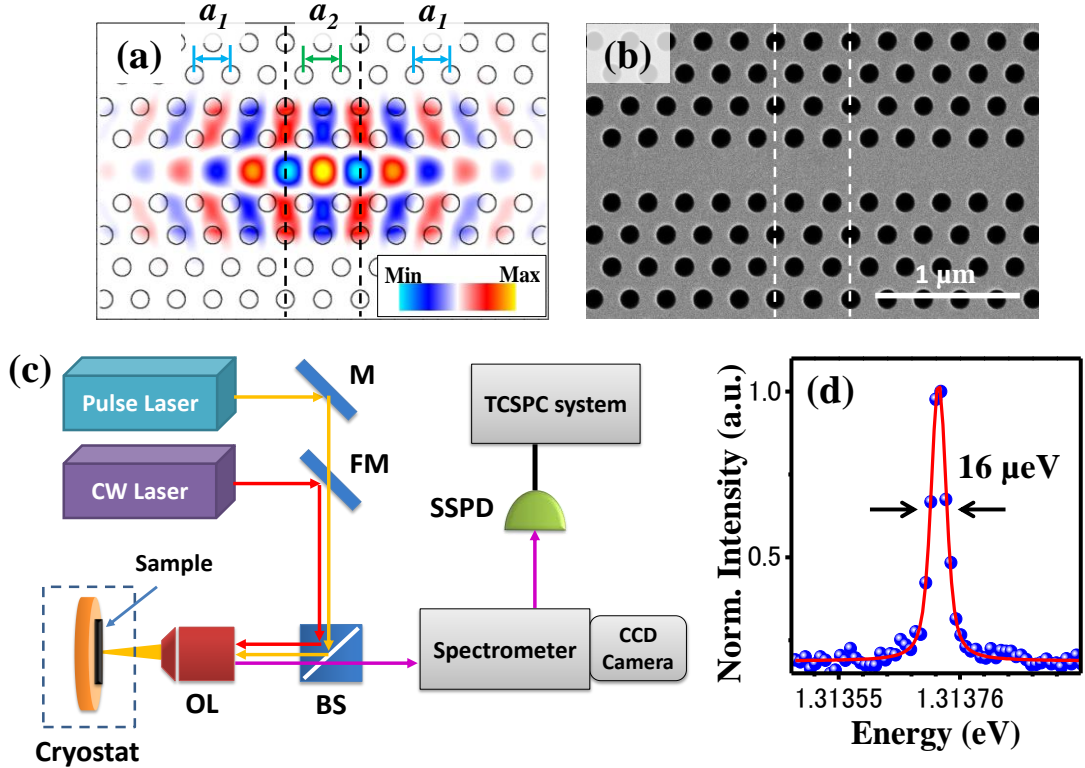

Fig. 1. (a) Simulated electric field profile of the investigated double-heterostructure cavity. The back dashed lines indicate the interfaces between the regular PhCs with a lattice constant of $a_1 = 252$ nm and the hetero region with $a_2 = 259.6$ nm. The width of the line defect waveguide is fixed to be $\sqrt{3}a_1$. (b) Top view SEM image of a fabricated cavity. The white dash lines indicate the hetero interfaces. (c) Schematic of the $\mu$PL measurement setup. M: mirror, FM: flip mirror, OL: objective lens, BS: beam splitter. (d) PL spectrum for the fundamental cavity mode measured under a far-detuned condition. Red solid line is of a fitting curve with a Voigt peak function, the Gaussian part of which corresponds to the spectrometer response.

Figure 2

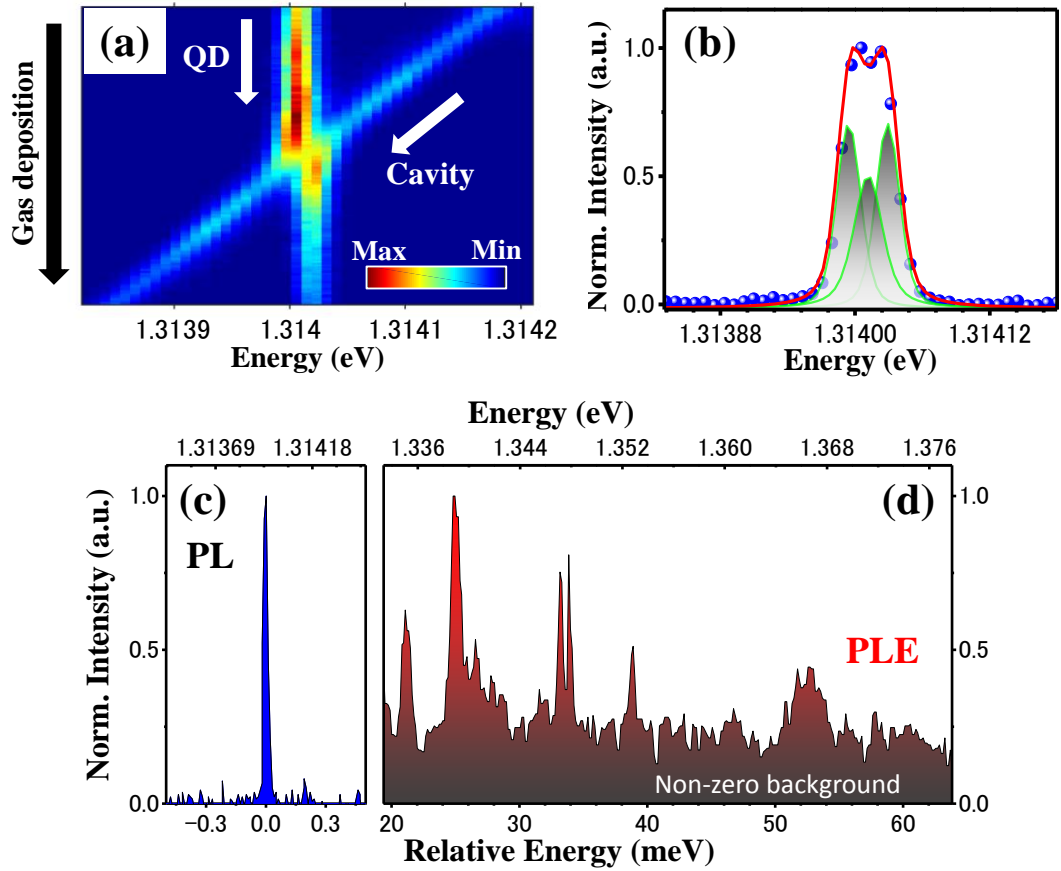

Fig. 2. (a) Color map of PL spectra measured under various QD-cavity detunings. (b) Vacuum Rabi spectrum measured at the QD-cavity resonance. The solid red line is of the fitting result by multiple Voigt peak functions. The light green lines show each component of the fitting function. The center peak between the two polariton peaks originates from the bare cavity emission. (c) PL spectrum of the investigated QD emitting at 1.31395 eV. (d) PLE spectrum for the QD PL peak in (c), exhibiting sharp PLE peaks and a broad non-zero background.

Figure 3

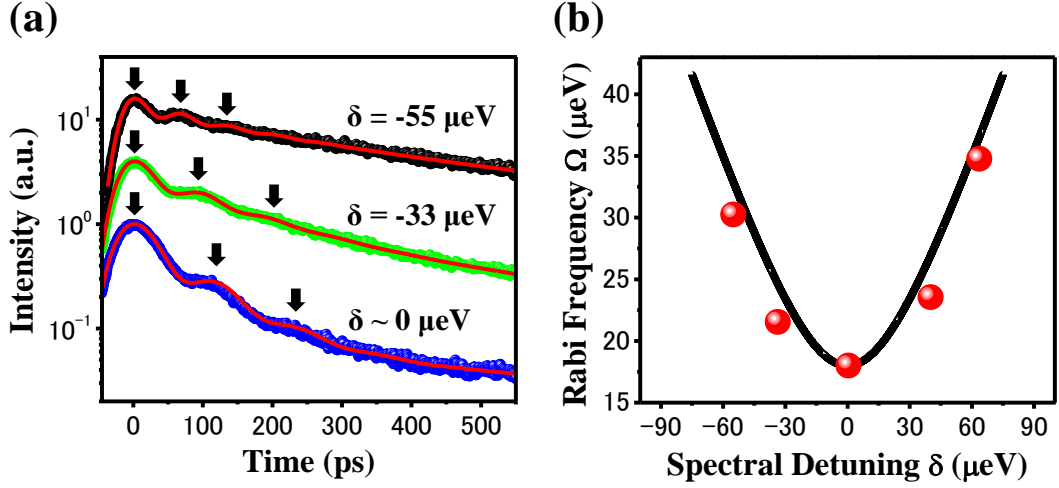

Fig. 3. (a) Time-resolved PL spectra measured at a laser excitation energy of 1.36564 eV under three-different QD-cavity spectral detunings of -55 $\mu$eV (black), -33 $\mu$eV (light green) and ~ 0 $\mu$eV (blue). Each decay curve is smoothed out by applying a moving average. The black arrows indicate peak positions of the oscillations. The red lines show the fitting curves. (b) Measured Rabi frequencies ($\Omega$) plotted as a function of QD-cavity spectral detuning $\delta$. The solid black line shows calculated $\Omega$s using a simple CQED model.

Figure 4

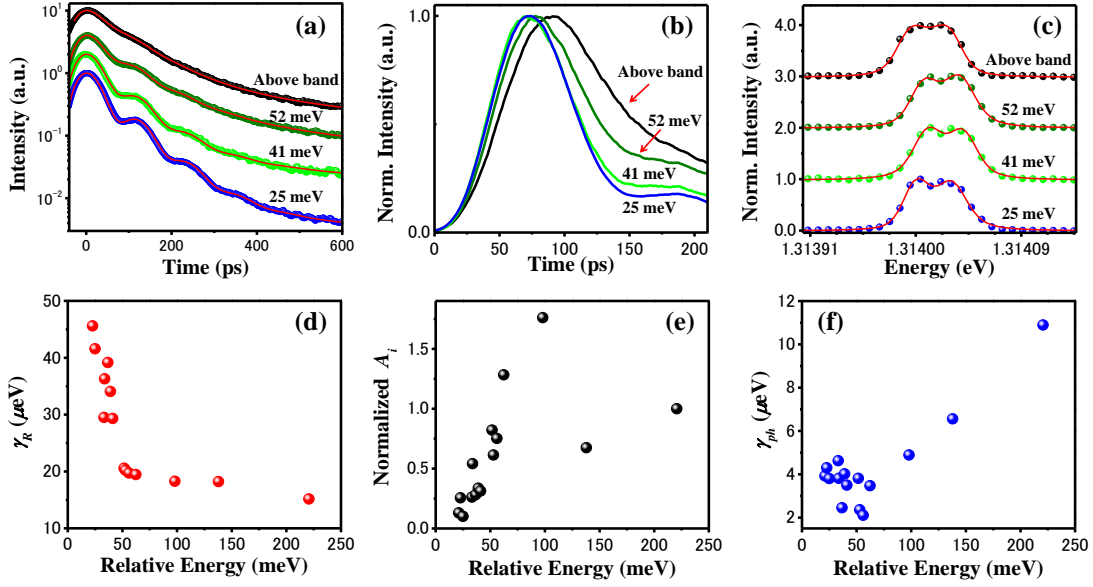

Fig. 4. (a) PL decay curves measured with varying the center energy of the pump laser. The QD and the cavity mode are under the resonance. The four curves are measured with various relative excitation energies: the black curve is for the excitation energy of 221 meV above the QD (above the GaAs bandgap), dark green for 52 meV, light green for 41 meV and blue for 25 meV. The solid red lines are the fitting curves. (b) Enlarged views for resolving the initial rises of time-domain PL curves corresponding to (a). (c) Corresponding vacuum Rabi spectra to (a) measured at the QD-cavity resonance. The red lines show the fitting curves. Slight shifts of the peak positions arise from the QD energy shifts under the different excitation conditions. (d) Extracted carrier relaxation rates $\gamma_R$s, (e) the intensities of the bare cavity emission $A_i$s and (f) pure dephasing rates $\gamma_{ph}$s plotted as a function of the relative excitation energies. $A_i$s are normalized to that measured under the above-bandgap excitation with the relative excitation energy of 221 meV. These data points were obtained by experiments performed under the conditions of keeping the same photon count rate on the SSPD by tuning the excitation power (between 100 nW and 140 $\mu$W).

Figure 5

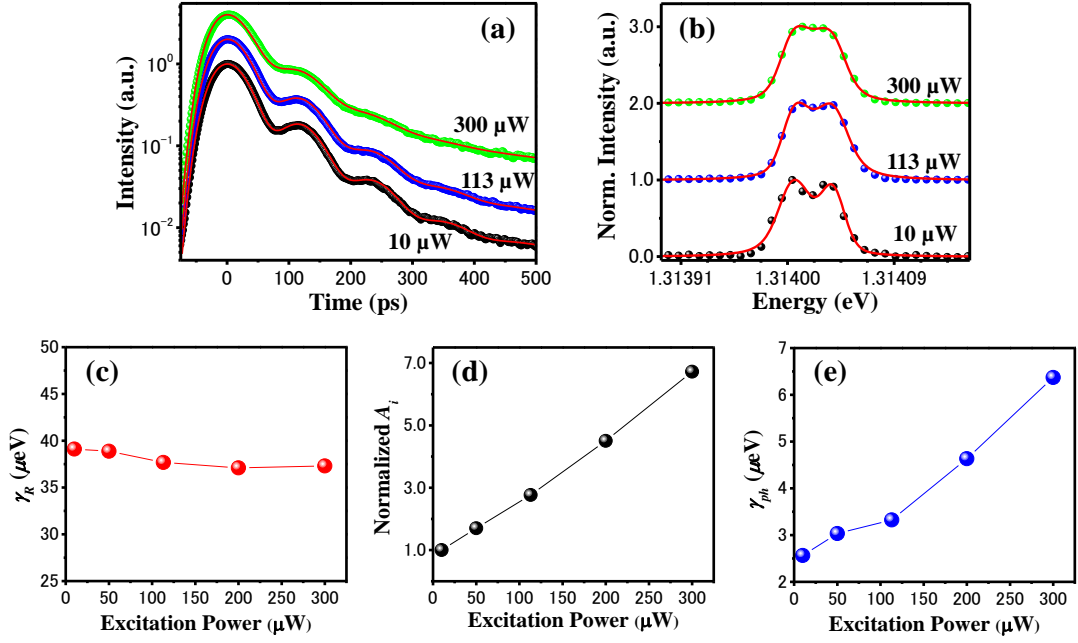

Fig. 5. (a) Time-resolved PL spectra measured at the QD-cavity resonance under the excitation power of 10 $\mu$W (black), 113 $\mu$W (red) and 300 $\mu$W (green). The red solid lines exhibit fitting results. (b) Corresponding PL spectra to (a), overlaid with fitting curves (red lines). (c) Extracted carrier relaxation rates $\gamma_R$s, (d) the intensities of the bare cavity emission $A_i$s, and (e) pure dephasing rates $\gamma_{ph}$s, plotted as a function of the excitation power. $A_i$s are plotted after normalized by that measured with a pump power of 10 $\mu$W